\documentclass[aps,pra,reprint,superscriptaddress,showpacs]{revtex4-1}
\usepackage{amssymb,graphicx,epsfig,threeparttable}
\usepackage{color}
\usepackage[colorlinks=true,linkcolor=blue]{hyperref}
\usepackage{ulem}

\newtheorem{defi}{Definition}
\newtheorem{prop}[defi]{Proposition}

\newtheorem{theo}[defi]{Theorem}

\begin{document}

\title{Hierarchy of non-Markovianity and $k$-divisibility phase diagram of quantum processes in open systems}
\author{Hong-Bin Chen}
\affiliation{Department of Physics, National Cheng Kung University, Tainan 701, Taiwan}
\author{Jiun-Yi Lien}
\affiliation{Department of Physics, National Cheng Kung University, Tainan 701, Taiwan}
\author{Guang-Yin Chen}
\affiliation{Department of Physics, National Chung Hsing University, Taichung 402, Taiwan}
\author{Yueh-Nan Chen}
\email{yuehnan@mail.ncku.edu.tw}
\affiliation{Department of Physics, National Cheng Kung University, Tainan 701, Taiwan}
\affiliation{Physics Division, National Center for Theoretical Sciences, Hsinchu 300, Taiwan}


\begin{abstract}
In recent years, many efforts have been devoted to the construction of a proper measure of quantum non-Markovianity. However, those proposed measures
are shown to be at variance in different situations. In this work, we utilize the theory of $k$-positive maps to generalize a hierarchy of
$k$ divisibility and develop a powerful tool, called $k$-divisibility phase diagram, to show the landscape of non-Markovianity. This allows one to study
different measures in a unified framework and provides a deeper insight into the nature of quantum non-Markovianity. By exploring the phase diagram with
several paradigms, we can explain the origin of the discrepancy between three frequently used measures and find the condition under which these measures
coincide with each other.
\end{abstract}

\maketitle

\section{INTRODUCTION}

Open quantum systems have attracted increasing attention due to their fundamental importance and applications in various fields. Generally speaking, one
has to deal with the interactions between the system and its environments, which may induce dissipation, decoherence, or rephasing. The environment usually
consists of a huge amount of degrees of freedom, and keeping track of them exactly is impossible. When dealing with the interactions, a typical approach is
to employ the Born-Markov approximation \cite{breuer_text_book}, leading to the celebrated Lindblad master equation \cite{gorini_lindbladform_jmp_1976,lindblad_comm_math_phys_1976}. Its solutions form a family of quantum dynamical semigroups. In an authentic physical
system, however, the dynamics is expected to deviate from the idealized Markovian evolution, in which the memory effects are essentially
ignored. In order to take the memory effects into account, many improved
techniques had been developed, such as path-integral formalisms \cite{leggett_rev,weiss_textbook,grabert_phys_rep,grifoni_phys_rep}, Monte Carlo
algorithms \cite{egger_pre}, hierarchy equations of motion (HEOM) \cite{tanimura_pra,tanimura_jpsj_1989,tanimura_jpsj_2006}, the
reaction-coordinate method \cite{nazir_pra_2014,garg_jcp_1985}, and
non-Markovian quantum master equations \cite{arimitsu_tc_jpsj_1980,mukamel_cop_pra_1978,silbey_comparison_cjp_2001}.

Although the notion of non-Markovianity has been used extensively, there is
no unique definition. Recently, many efforts have been devoted to the
construction of an appropriate measure of quantum non-Markovianity \cite{plenio_review_rep_prog_phys_2014}. Most of them are based on continuously
monitoring the time variation of certain quantities of interest, such as
trace-distance \cite{BLP_measure_prl_2009}, entanglement \cite{RHP_measure_prl_2010}, mutual information \cite{LFS_measure_pra_2012},
channel capacity \cite{BCM_measure_sci_rep_2014}, and the set of accessible
states \cite{LPP_measure_pra_2013}. When these quantities decrease
monotonically in time, the system undergoes a Markovian process. On the
other hand, whenever the revival of these quantities is detected, the
corresponding measure of non-Markovianity can be constructed according to an
optimal amount of the revival.

In the non-Markovianity measure proposed by Breuer, Lane, and Piilo (BLP) %
\cite{BLP_measure_prl_2009}, the authors focus on the trace-distance $%
\mathrm{D}\left(\rho_{1,2};t\right)=\Vert\mathcal{E}_{t,0}\left(\rho_1-%
\rho_2\right)\Vert_{1}/2$ of a pair of arbitrary initial states $\rho_{1}$
and $\rho_{2}$, where $\mathcal{E}_{t,0}$ is a completely positive (CP) and
trace-preserving (TP) quantum process and $\Vert A\Vert_{1}=\text{Tr}\sqrt{%
A^{\dagger}A}$ denotes the trace norm of a matrix $A$. They define the
time-varying rate $\sigma\left(\rho_{1,2};t\right)=\frac{\partial}{\partial t%
}\mathrm{D}\left(\rho_{1,2};t\right)$ and interpret $\sigma\left(%
\rho_{1,2};t\right)$ as an information flow. Then $\sigma\left(\rho_{1},%
\rho_{2};t\right)>0$ witnesses a backflow of information from the
environment into the system which increases the distinguishability of $%
\rho_{1}$ and $\rho_{2}$ and indicates the non-Markovian character of the
process $\mathcal{E}_{t,0}$.

Focusing on the information flux, the measure proposed by Luo, Fu, and Song
(LFS) \cite{LFS_measure_pra_2012} is characterized by an
information-theoretic quantity. The authors consider a maximally-entangled
system-ancilla pair and monitor the time evolution of the quantum mutual
information $\mathrm{I}\left( \mathrm{S}:\mathrm{A}\right) =S\left( \rho ^{%
\mathrm{S}}\right) +S\left( \rho ^{\mathrm{A}}\right) -S\left( \rho ^{%
\mathrm{SA}}\right) $, where $S\left( \rho ^{\mathrm{S}}\right) =-\mathrm{Tr}%
\rho ^{\mathrm{S}}\log _{2}\rho ^{\mathrm{S}}$ is the von Neumann entropy
and $\rho ^{\mathrm{S}}=\mathrm{Tr}_{\mathrm{A}}\rho ^{\mathrm{SA}}$ is the
reduced density matrix of the system. The mutual information quantifies the
correlation of a bipartite system and never increases under a local CPTP
quantum process. Hence, the revival of mutual information indicates that the
lost information flows back and the correlation between the system-ancilla
pair is recovered.

The underlying origin of the revival of these quantities is the divisibility
of the processes \cite{wolf_cmp_2008,wolf_prl_2008}. A CPTP quantum process $%
\mathcal{E}_{t,0}$ is said to be CP divisible if, $\forall ~t,\tau >0$,
there exists a \textit{complement process} $\Lambda _{t+\tau ,t}$ which is
also CPTP and satisfies the composition law $\mathcal{E}_{t+\tau ,0}=\Lambda
_{t+\tau ,t}\circ \mathcal{E}_{t,0}$. A CPTP process $\mathcal{E}_{t,0}$ is
said to be Markovian if and only if it is CP divisible. Hence, the measure of
non-Markovianity proposed by Rivas, Huelga, and Plenio (RHP) \cite%
{RHP_measure_prl_2010} is the degree of a process deviating from being
CP divisible, i.e., $g(t)=\lim_{\epsilon \rightarrow 0^{+}}\left[ \Vert
\left( \mathcal{I}_{\mathrm{anc}}\otimes \Lambda _{t+\epsilon ,t}\right)
\left( |\Psi \rangle \langle \Psi |\right) \Vert _{1}-1\right] /\epsilon $,
where $|\Psi \rangle $ denotes a maximally entangled state between the
system and a copy of well-isolated ancilla possessing the same degrees of
freedom of the system and $\mathcal{I}_{\mathrm{anc}}$ is the identity
process acting on the ancillary degree of freedoms. Due to the Choi-Jamio{\l
}kowski isomorphism \cite{jamiolkowski_rmp_1972,choi_laa_1975}, the
complement process $\Lambda _{t+\epsilon ,t}$ is CPTP if and only if $g(t)=0$%
, $\forall ~t>0$, namely, $\mathcal{E}_{t,0}$ is CP divisible and Markovian.

Apart from some special cases in which only a \textit{single} decoherence
channel is present \cite{LFS_measure_pra_2012,chinapeople_comparison_pra_2011,haikka_comparison_pra_2012},
many comparative studies \cite{chruscinski_comparison_pra_2011,apollaro_comparison_pra_2014,addis_comparison_pra_2014}
show that these measures are essentially at variance, especially when the
process consists of multiple decoherence channels.

In view of the discrepancies among distinct measures, the following
questions naturally arise. What is the underlying reason that makes these
measures at variance and under what circumstance do these measures coincide
with each other? Is it possible to construct a better measure when all the
measures fail to work? To address these questions, we adopt the concept of $k$ \textit{divisibility} \cite{chruscinski_k_divi_prl_2014}, a natural
generalization of CP divisibility, to develop the $k$ \textit{divisibility
phase diagram}, which can show the landscape of non-Markovianity and allow
us to study different measures in a unified framework. Furthermore, we can
acquire more fine-grained insights into the nature of quantum
non-Markovianity by exploring the phase diagram.

\section{$k$ POSITIVILITY}

We first explicitly introduce the notion of Choi-Jamio{\l }kowski
isomorphism \cite{jamiolkowski_rmp_1972,choi_laa_1975}. Suppose that $\mathcal{A}$ is a C$^{\ast }$ algebra of linear operators on the
$n$-dimensional Hilbert space $\mathcal{H}_{n}$ and $\mathfrak{L}\left(\mathcal{A},\mathcal{A}\right) $ denotes the set of linear maps from
$\mathcal{A}$ to $\mathcal{A}$. Assuming that $\left\{ E_{i}\right\} $ is an
orthonormal basis for $\mathcal{A}$ with $\mathrm{Tr}E_{j}^{\dagger}E_{i}=\delta _{i,j}$, the Choi-Jamio{\l }kowski isomorphism can be defined
as a linear map $\mathfrak{J}:\mathfrak{L}\left(\mathcal{A},\mathcal{A}\right)\rightarrow\mathcal{A}\otimes\mathcal{A}$ by
\begin{equation}
\mathfrak{J}\left(\mathcal{E}\right)=\sum_{i}^{n} E_{i}\otimes\mathcal{E}\left(E_{i}\right).
\end{equation}

Let $\mathcal{A}^{+}$ be the subset of positive elements; a TP map $\mathcal{E}\in\mathfrak{L}\left(\mathcal{A},\mathcal{A}\right)$ is said to be
positivity preserving (PP) if $\mathcal{E}\left(\mathcal{A}^{+}\right)\subset\mathcal{A}^{+}$. However, a quantum system may have nonclassical
correlations (e.g., entanglement) with some other ancillary degrees of
freedom. To ensure that one can always obtain a legitimate quantum state in
the presence the such nonclassical correlations, the notion of PP must be
generalized to a series of $k$ positivity: A TP map $\mathcal{E}$ is said to
be $k$ positive if $\mathcal{I}_{k}\otimes\mathcal{E}:\mathcal{M}_{k}\otimes\mathcal{A}\rightarrow \mathcal{M}_{k}\otimes\mathcal{A}$ is PP and CP if
$\mathcal{E}$ is $k$ positive for all positive integers $k$, where $\mathcal{I}_{k}$ is the identity map acting on the $k\times k$ matrix algebra
$\mathcal{M}_{k}$. Although the structure of CP maps have been studied
thoroughly \cite{jamiolkowski_rmp_1972,choi_laa_1975}, there is still no
efficient criterion for determining whether a map is $k$ positive or not \cite{randade_k_posi_criterion_osid_2007,skowronek_k_posi_criterion_jmp_2009,hou_k_posi_criterion_laa_2015}.

For textual completeness, here we briefly introduce some criteria for $k$ positivity and the famous Choi's criterion for CP. Let
$\left\{|j\rangle\right\}$ and $\left\{|\psi_j\rangle\right\}$ be the standard basis for
$\mathbb{C}^{k}$ and a set of arbitrary $k$ mutually orthonormal vectors in $\mathcal{H}_n$, respectively. Then it is easy to see that
$|\psi\rangle=k^{-1/2}\sum_{j=1}^{k}|j\rangle\otimes|\psi_j\rangle$ is a
vector of Schmidt rank $k$ and $P=|\psi\rangle\langle\psi|$ is a rank-$1$
projection operator in $\mathcal{M}_k\otimes\mathcal{A}$. Then the $k$ positivity of a linear map can be verified via the following criteria:

\begin{prop}[criteria for $k$ positivity]\label{prop_criteria_for_k_positivity_1}
For a linear map $\mathcal{E}\in\mathfrak{L}\left(\mathcal{A},\mathcal{A}\right)$, the following are equivalent.
\begin{enumerate}
\item $\mathcal{E}$ is $k$ positive.
\item $\left(\mathcal{I}_{k}\otimes\mathcal{E}\right)\left(P\right)$ is a positive semidefinite matrix in $\mathcal{M}_k\otimes\mathcal{A}$ for all rank $1$ projection operator $P\in\mathcal{M}_k\otimes\mathcal{A}$.
\item $\sum_{i,j=1}^{k} |i\rangle\langle j|\otimes\mathcal{E}\left(|\psi_{i}\rangle\langle\psi_{j}|\right)$ is a positive semidefinite matrix
in $\mathcal{M}_k\otimes\mathcal{A}$.
\end{enumerate}
\end{prop}

The proof of the above proposition are quite long and beyond the scope of this work. For details, please see
Refs.~\cite{randade_k_posi_criterion_osid_2007,skowronek_k_posi_criterion_jmp_2009,hou_k_posi_criterion_laa_2015} and the references therein.

From the definition of $k$ positivity, one notices that PP and CP are special cases for $k=1$ and $n$, respectively.
However, to perform the criteria for positivity or $k$ positivity is usually a tedious task since the validity should be held for all elements in the corresponding space.
Fortunately, Choi proved an efficient way to verify the CP property \cite{choi_laa_1975}.

\begin{theo}[Choi's theorem on CP map]\label{theo_criterion_for_cp}
A TP map $\mathcal{E}\in\mathfrak{L}\left(\mathcal{A},\mathcal{A}\right)$ is CP if and only if its corresponding Choi-Jamio{\l}kowski matrix $\mathfrak{J}\left(\mathcal{E}\right)$
is positive semidefinite.
\end{theo}

This theorem is important since it greatly simplifies the task for verifying the CP property. One only needs to evaluate the eigenvalues of $\mathfrak{J}\left(\mathcal{E}\right)$
instead of the formidable task performed on all elements in each corresponding space.

\section{$k$ DIVISIBILITY AND $\mathcal{PD}_k$ PARTITION}

Having the notion of $k$-positive maps, we can generalize CP divisibility to a hierarchy of $k$ divisibility: A CPTP quantum process $\mathcal{E}_{t,0}$ is said to be $k$ divisible
if, $\forall~t,\tau > 0$, the complement process
\begin{equation}
\Lambda_{t+\tau,t}=\mathcal{E}_{t+\tau,0}\circ\left[\mathcal{E}_{t,0}\right]^{-1}
\end{equation}
is $k$ positive. Consequently, $n$ divisibility is equivalent to CP divisibility and
$\mathcal{E}_{t,0}$ is $0$-divisible if $\Lambda_{t+\tau,t}$ is not a PP map. Introducing a family of sets $\mathcal{D}_{k}$ containing processes $\mathcal{E}_{t,0}$
with divisibility less than $k$, one has a chain of inclusions,
\begin{equation}
\mathcal{D}_{0}\subset\mathcal{D}_{1}\subset\cdots\subset\mathcal{D}_{n-1}\subset\mathcal{D}_{n},
\label{k_divi_inclusion_chain}
\end{equation}
where $\mathcal{D}_{n}$ consists of all quantum processes, Markovian or non-Markovian, and $\mathcal{D}_{0}$ consists of $0$-divisible processes, which is called
\textit{essentially non-Markovian} by Chru\'{s}ci\'{n}ski \textit{et al.}~\cite{chruscinski_k_divi_prl_2014}. Now we propose to define the sets of proper $k$ divisibility
$\mathcal{PD}_{k}=\mathcal{D}_{k}-\mathcal{D}_{k-1}$; then $\mathcal{PD}_{n}=\mathcal{D}_{n}-\mathcal{D}_{n-1}$ consists of processes which are exactly $n$ divisible, i.e.,
Markovian processes. Thus, the inclusion chain in Eq.~(\ref{k_divi_inclusion_chain}) can be rewritten into a partition of $\mathcal{D}_{n}$ in terms of $\mathcal{PD}_{k}$
\begin{equation}
\mathcal{D}_{n}=\bigcup_{k=0}^{n}\mathcal{PD}_{k}.
\label{k_divi_partition}
\end{equation}

Now we can summarize the above discussions with the algorithm to the partition in Eq.~(\ref{k_divi_partition}).
Let $\mathcal{E}_{t,0}\in\mathfrak{L}\left(\mathcal{A},\mathcal{A}\right)$
be a CPTP process and $\mathfrak{J}\left(\mathcal{E}_{t,0}\right)$ its corresponding
Choi-Jamio{\l}kowski matrix. One can use the following procedures to
determine the $\mathcal{PD}_{k}$ partition in Eq.~(\ref{k_divi_partition}).

\begin{enumerate}
\item Using Choi's criterion for CP in Theorem~\ref{theo_criterion_for_cp}, one can make sure the CP property of $\mathcal{E}_{t,0}$ or under which circumstance the process $\mathcal{E}_{t,0}$ is CP.
\item Calculate the complement process $\Lambda_{t+\tau,t}=\mathcal{E}_{t+\tau,0}\circ\left[\mathcal{E}_{t,0}\right]^{-1}$.
\item Applying Choi's criterion for CP in Theorem~\ref{theo_criterion_for_cp} to $\Lambda_{t+\tau,t}$, one can verify under which circumstance $\Lambda_{t+\tau,t}$ will be CP $\forall~t,\tau\geq 0$. This leads to the region of $\mathcal{PD}_n$.
\item Applying the criteria in Proposition~\ref{prop_criteria_for_k_positivity_1} to $\Lambda_{t+\tau,t}$, one can verify under which circumstance
$\Lambda_{t+\tau,t}$ will be $(n-1)$ positive. This leads the the region of $\mathcal{PD}_{n-1}$.
\item Continue step $\mathrm{4.}$ until one obtains the region of $\mathcal{PD}_{0}$.
\end{enumerate}

To make the partition in Eq.~(\ref{k_divi_partition}) intuitive, in the following, we apply the algorithm to several paradigms and show the explicit visualization by means of
a $k$-divisibility phase diagram. This helps us to address the questions proposed in the Introduction and reveals more of the nature of quantum non-Markovianity. More detailed
expressions can be found in the Appendix. Additionally, a similar partition has been reported recently in an all optical setup \cite{bernardes_exp_arxiv_2015}.

\section{$k$-DIVISIBILITY PHASE DIAGRAM}

\subsection{Pauli dephasing channel}

\begin{figure}[th]
\includegraphics[width=\columnwidth]{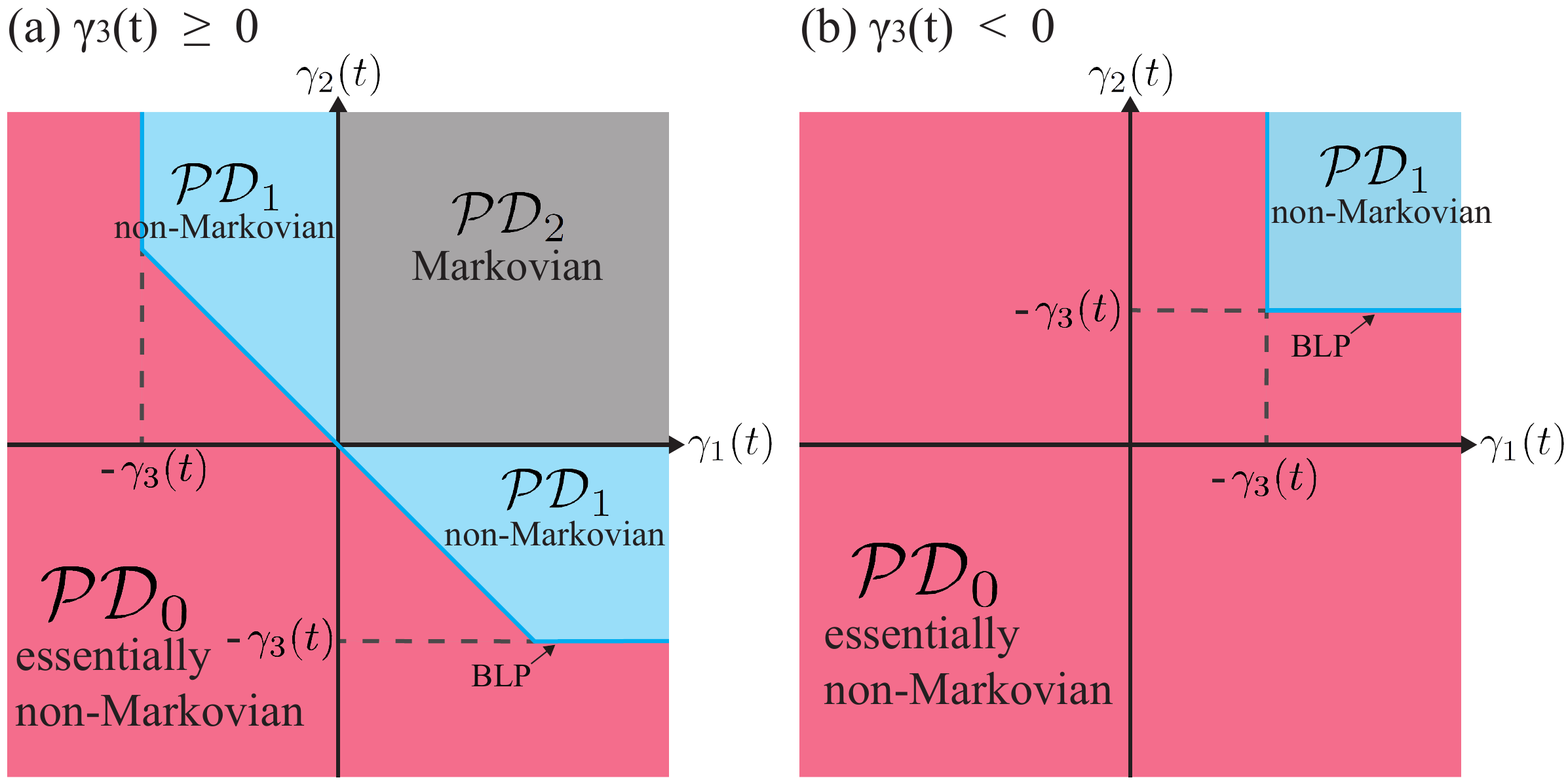}
\caption{(Color online) The $k$-divisibility phase diagram for the Pauli dephasing channel in three-dimensional $\gamma_{j}(t)$ space with (a) $\gamma_{3}(t)\geq 0$ and
(b) $\gamma_{3}(t)<0$. The $\mathcal{PD}_{2}$ (gray) is for the Markovian processes with nonnegative rates $\gamma_{j}(t)\geq 0$ confined in the first octant, $\mathcal{PD}_{1}$ (blue)
is for the non-Markovian processes with rates satisfying conditions in Eqs.~(\ref{condi_PD1_pauli_dephasing}), and $\mathcal{PD}_{0}$ (red) is for the essentially non-Markovian processes
away from the above regions. The upper bound of the BLP measure coincides with the border between $\mathcal{PD}_{1}$ and $\mathcal{PD}_{0}$.}
\label{Fig_exam_1}
\end{figure}

First, we consider a master equation for a qubit in the standard Lindblad form
\begin{equation}
\frac{\partial}{\partial t}\rho=\mathcal{L}_{t}\rho=\frac{1}{2}\sum_{j=1}^{3}\gamma_{j}(t)\left(\hat{\sigma}_{j}\rho\hat{\sigma}_{j}-\rho\right),
\label{lindblad_form_pauli_dephasing}
\end{equation}
where $\hat{\sigma}_{j}$ are the Pauli matrices. It is well known that the necessary and sufficient conditions for its corresponding process
$\mathcal{E}_{t}=\mathcal{T}\exp\left[\int_{0}^{t}\mathcal{L}_{\tau}d\tau\right]\in\mathcal{PD}_{2}$ (being Markovian) are
\begin{equation}
\gamma_{1}(t)\geq 0,~\gamma_{2}(t)\geq 0,~\gamma_{3}(t)\geq 0.
\label{condi_PD2_pauli_dephasing}
\end{equation}
On the other hand, the conditions for $\mathcal{E}_{t}\in\mathcal{PD}_{1}$ are weaker than those of Eqs.~(\ref{condi_PD2_pauli_dephasing}) \cite{chruscinski_k_divi_prl_2014}; i.e.,
\begin{eqnarray}
\gamma_{1}(t)+\gamma_{2}(t)\geq 0,\nonumber\\
\gamma_{2}(t)+\gamma_{3}(t)\geq 0,\nonumber\\
\gamma_{3}(t)+\gamma_{1}(t)\geq 0.
\label{condi_PD1_pauli_dephasing}
\end{eqnarray}
Although the region of Eqs.~(\ref{condi_PD2_pauli_dephasing}) is enclosed by that of Eqs.~(\ref{condi_PD1_pauli_dephasing}), $\mathcal{PD}_{2}$ should be excluded from $\mathcal{PD}_{1}$ by definition. The above conditions can be depicted in a three-dimensional $\gamma_{j}(t)$ space for a clear comprehension. In Fig.~\ref{Fig_exam_1}, we show the half space with (a)
$\gamma_{3}(t)\geq 0$ and (b) $\gamma_{3}(t)<0$. The first octant consists of all Markovian processes satisfying Eqs.~(\ref{condi_PD2_pauli_dephasing}) and therefore lies in the region
of $\mathcal{PD}_{2}$ (gray). The $\mathcal{PD}_{1}$ (blue) consists of non-Markovian processes satisfying Eqs.~(\ref{condi_PD1_pauli_dephasing}) surrounding the region of
$\mathcal{PD}_{2}$, and the $\mathcal{PD}_{0}$ (red) consists of essentially non-Markovian ones. 
Furthermore, it has been shown that whenever
$\mathcal{E}_{t}\in\mathcal{PD}_{1}$, $\sigma\left(\rho_{1},\rho_{2};t\right)\leq 0$ is always fulfilled \cite{chruscinski_k_divi_prl_2014,chruscinski_phys_lett_a_2013}. Namely, BLP
measure can only detect the non-Markovianity in the region of $\mathcal{PD}_{0}$ and be blind to the non-Markovianity in the region of $\mathcal{PD}_{1}$. The underlying reason for this
weakness lies in that the definition of $\sigma\left(\rho_{1},\rho_{2};t\right)$ does not take the full advantage of CP divisibility.

To further show the powerfulness of the divisibility phase diagrams, we consider the \textit{eternal non-Markovian} process proposed by Hall \textit{et al.} \cite{hall_pra_2014}
with decay rates
\begin{equation}
\gamma_{1}(t)=\gamma_{2}(t)=1,~\gamma_{3}(t)=-\tanh t.
\end{equation}
It is clear that these decay rates lie exactly in the region of $\mathcal{PD}_{1}$; hence, the corresponding process is non-Markovian and can never be detected by BLP measure.
Along the same line, one can easily construct more eternal non-Markovian processes such as
\begin{equation}
\gamma_{1}(t)=1,~\gamma_{2}(t)=-\gamma_{3}(t)=\sin t.
\end{equation}

\subsection{CNOT gate}

\begin{figure}[th]
\includegraphics[width=\columnwidth]{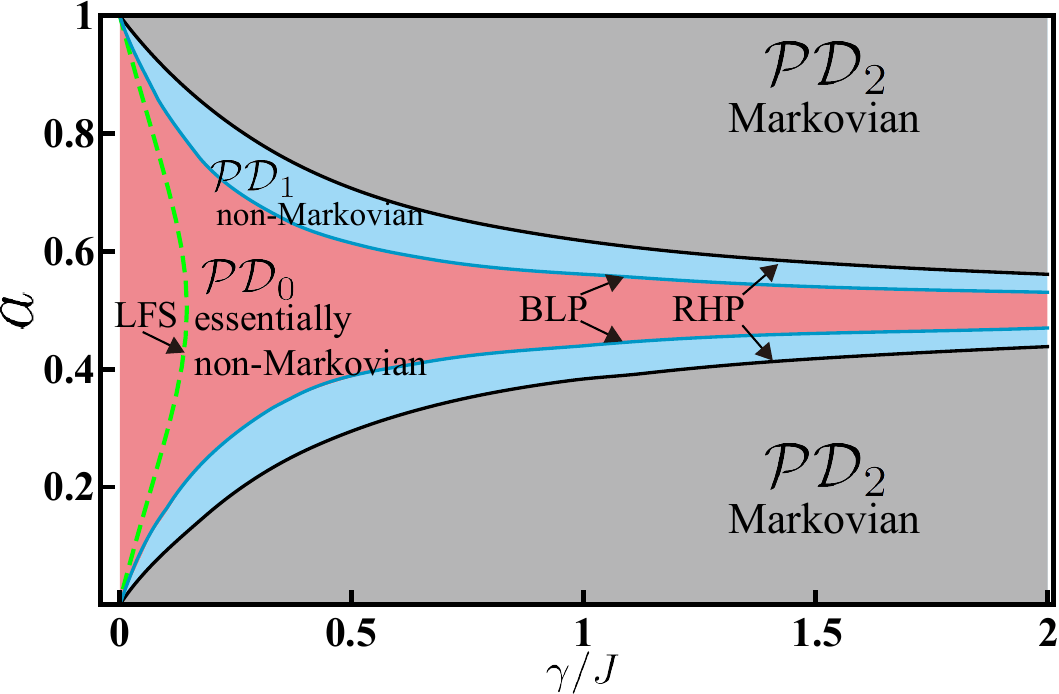}
\caption{(Color online)
The $k$-divisibility phase diagram for the T qubit controlled by the C qubit via a controlled-NOT (CNOT) gate. Besides the CNOT gate, the T qubit undergoes an isotropic depolarizing
channel. The C qubit is a mixture and has population $a$ on $|1\rangle\langle 1|$. For $a=1$ or $0$, the C-qubit is pure and has no information that can flow into the T qubit. Hence,
the T qubit is Markovian ($\mathcal{PD}_2$). As $a$ approaches $0.5$, the T qubit shows a transition from $\mathcal{PD}_2$ to $\mathcal{PD}_0$. In this case, the RHP measure is shown
to be perfect and can thoroughly detect the non-Markovianity in both regions of $\mathcal{PD}_0$ and $\mathcal{PD}_1$. The BLP measure can detect the non-Markovianity in
$\mathcal{PD}_0$, whereas the LFS measure can only work on the left side of the green dashed curve.}
\label{Fig_exam_2}
\end{figure}

The second example considered is a pair of qubits coupled with each other via a controlled-NOT (CNOT) gate, which is a fundamental building block in the theory of quantum information.
Assume that the control qubit (C qubit) is a mixture $\rho^{\mathrm{C}}=a|1^\mathrm{C}\rangle\langle 1^\mathrm{C}|+(1-a)|0^\mathrm{C}\rangle\langle 0^\mathrm{C}|$ and the qubit pair
has no initial interqubit correlation. Their interaction Hamiltonian can be written as
\begin{equation}
\widehat{H}_{\mathrm{CT}}=\frac{J}{2}\left[|1^\mathrm{C}\rangle\langle 1^\mathrm{C}|\otimes\hat{\sigma}_x+|0^\mathrm{C}\rangle\langle 0^\mathrm{C}|\otimes\hat{\mathcal{I}}\right].
\end{equation}
Besides, the target qubit (T qubit) undergoes a noisy channel which is described by isotropic depolarizing channel with
\begin{equation}
\gamma_{1}(t)=\gamma_{2}(t)=\gamma_{3}(t)=\gamma.
\end{equation}
The parameter $a$ has prominent influences on the dynamics of the T qubit. In Fig.~\ref{Fig_exam_2}, we explore the $k$-divisibility phase diagram for the T qubit in the $\gamma-a$
plane. When $a=1$ or $0$, the C qubit is a pure sate in $|1^\mathrm{C}\rangle$ or $|0^\mathrm{C}\rangle$, respectively, and contains no information. Consequently, the T qubit can only
decohere Markovianly. As $a$ gradually approaches $0.5$, the C qubit becomes more uncertain and contains more information. The information flow into the T qubit resultantly dominates
over the Markovian decoherence channel and the T qubit behaves non-Markovianly. As a result, the dynamics of the T qubit shows a transition from $\mathcal{PD}_2$ to $\mathcal{PD}_0$
with $a$ approaching $0.5$ or reducing $\gamma$ in Fig.~\ref{Fig_exam_2}.

Moreover, we find out that the upper bound of RHP measure coincides with the border between $\mathcal{PD}_1$ and $\mathcal{PD}_2$. This means that, in this case, the RHP measure is a
perfect measure which can thoroughly detect the non-Markovianity in both regions of $\mathcal{PD}_0$ and $\mathcal{PD}_1$. On the other hand, the upper bound of BLP measure also
coincides with the border between $\mathcal{PD}_0$ and $\mathcal{PD}_1$, and the reason is similar to that in the previous example. Besides, the upper bound of LFS measure is indicated
by the green dashed curve in Fig.~\ref{Fig_exam_2}. This illustrates the fact that the LFS measure can only work on the left side of the green dashed curve and is, in general, much
weaker than BLP measure.

We attribute this weakness to the overcrucial requirement of mutual
information. The BLP measure ensures that the dynamics in $\mathcal{PD}_{0}$
indeed undergoes information backflow. This backflowing information,
however, is alien to the information-theoretic quantities such as von
Neumann entropy and mutual information. BLP refer the time varying of
trace distance as information flow from the angle of the distinguishability
of a pair of states, while increasing the distinguishability of a pair of
states does not guarantee the increasing of the correlation with ancilla.
This makes the LFS measure underestimate the information backflow and fail
to detect the non-Markovianity. A similar conclusion can be found in Ref.~\cite{apollaro_comparison_pra_2014} and leads to the chain of inclusions
\begin{equation}
\mathrm{LFS}\subseteq\mathrm{BLP}\subseteq\mathrm{RHP}.
\end{equation}

\subsection{Amplitude-damping channel}

\begin{figure}[h]
\includegraphics[width=\columnwidth]{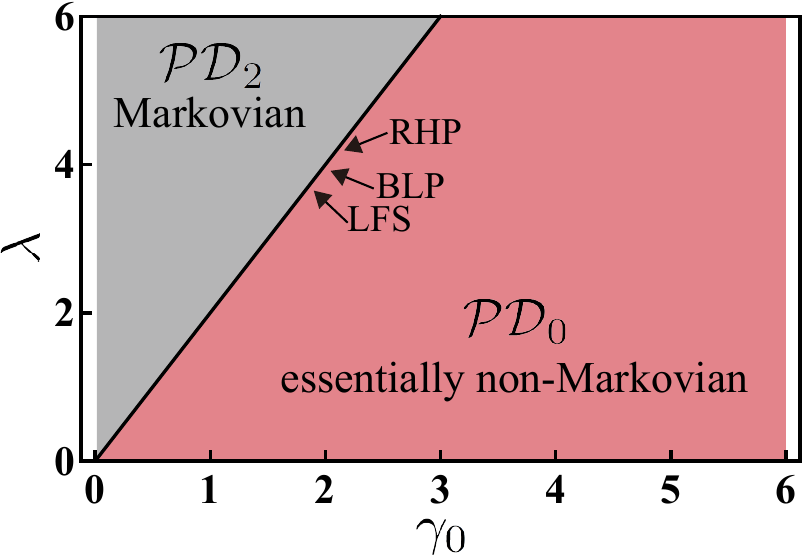}
\caption{(Color online)
The $k$-divisibility phase diagram for a single amplitude-damping channel with the Lorentzian spectral density. The criteria for both $\mathcal{PD}_{0}$ and $\mathcal{PD}_{1}$ are
exactly the same ($\gamma_{0}>\lambda/2$), and the phase diagram shows the degeneracy explicitly. In this model, both the BLP and the LSF measures can be as perfect as RHP due to
the degeneracy.}
\label{Fig_exam_3}
\end{figure}

In the following, we proceed to address the question of the circumstance under which these measures can be perfect, as shown in
Refs.~\cite{LFS_measure_pra_2012,chinapeople_comparison_pra_2011,breuer_pra_2010}.
We consider the qubit dynamics subjected to a amplitude-damping channel
\begin{equation}
\frac{\partial}{\partial t}\rho=\gamma(t)\left(\hat{\sigma}_{-}\rho\hat{\sigma}_{+}-\frac{1}{2}\left\{\hat{\sigma}_{+}\hat{\sigma}_{-},\rho\right\}\right).
\end{equation}
The time-varying rate $\gamma(t)$ is determined by the spectral density function $J(\omega)$ of the environment. In this example, we consider the Lorentzian spectral density
function $J(\omega)=\gamma_{0}\lambda^{2}/\left[\pi\left(\omega^{2}+\lambda^{2}\right)\right]$. In Fig.~\ref{Fig_exam_3}, we depict the $k$-divisibility phase diagram in the
$\gamma_{0}-\lambda$ plane, and the criteria for $\mathcal{PD}_{2}$ and $\mathcal{PD}_1$ are exactly the same. Consequently, the $\gamma_{0}-\lambda$ plane is divided into only
two parts, $\mathcal{PD}_2$ and $\mathcal{PD}_0$, and the degeneracy of $\mathcal{PD}_1$ can be shown explicitly by the divisibility phase diagram for such dynamics possessing
only one decoherence channel. As expected, the BLP measure can detect all non-Markovianity in the region of $\mathcal{PD}_{0}$ with $\gamma _{0}>\lambda /2$. Due to the degeneracy,
both the BLP and the LFS measures can detect all non-Markovianity in the $\gamma_{0}-\lambda$ plane and are equivalent to the RHP measure for this model. This is in line with the
conclusions of Refs.~\cite{LFS_measure_pra_2012,chinapeople_comparison_pra_2011,breuer_pra_2010}.

Additionally, a similar conclusion can be drawn in a special case of the first example, where the qubit system is subject to only one of the three dephasing channels, e.g.,
$\gamma_{1}(t)=\gamma_{2}(t)=0$ and $\gamma_{3}(t)\neq 0$. The border of $\mathcal{PD}_1$ meets that of $\mathcal{PD}_2$ at the origin in Fig.~\ref{Fig_exam_1}, and the degeneracy
occurs alone the $\gamma_{3}$ axis. The BLP measure again works perfectly as the RHP measure does, as shown in Ref.~\cite{chinapeople_comparison_pra_2011}.

\subsection{Superradiance}

\begin{figure}[h]
\includegraphics[width=\columnwidth]{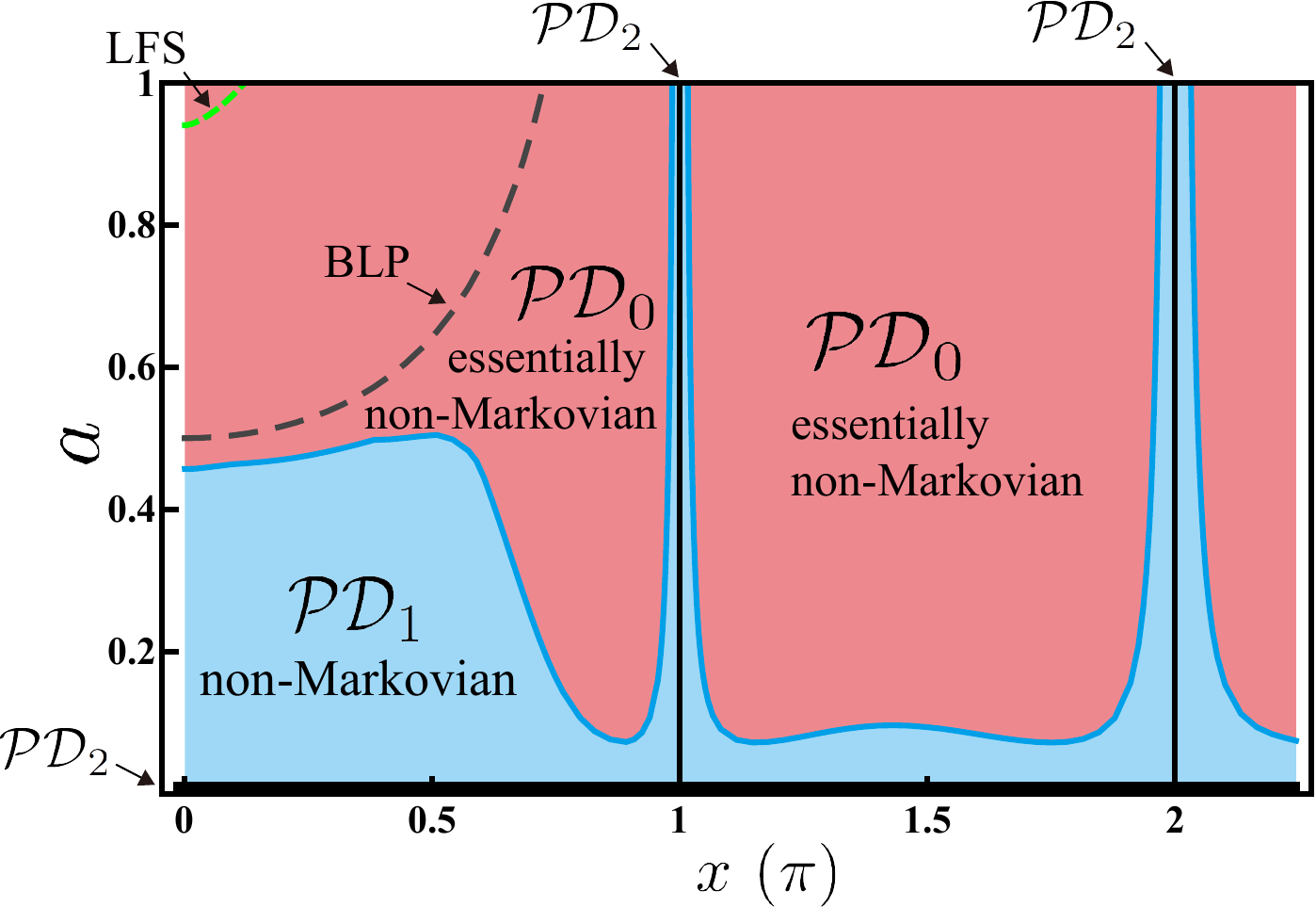}
\caption{(Color online) The $k$-divisibility phase diagram for a pair of atoms with the superradiant process, where $x$ is defined as the interatomic distance $d$ multiplied by the
wave number $q$. The environment atom has an initial population $a$ in its excited state, which is related to the rate of energy feedback to the system atom. For $a=0$ or $x=n\pi$,
the system is Markovian ($\mathcal{PD}_{2}$). The region surrounding $\mathcal{PD}_{2}$ with small $a$ or $x$ close to $n\pi$ is dominated by $\mathcal{PD}_{1}$. When $a$ increases,
the system atom becomes more non-Markovian and is dominated by $\mathcal{PD}_{0}$. The non-Markovianity reduces to Markovianity and appears periodically when increasing $x$. In this
case, the BLP measure can only partially detect the non-Markovianity in the region of $\mathcal{PD}_{0}$ (the region above the dashed curve), and the LFS measure can even only work
in the small corner above the green dashed curve.}
\label{Fig_exam_4}
\end{figure}

Apart from these theoretical models described by the standard Lindblad form, we finally consider the superradiant phenomenon in a two-atom system which can be implemented experimentally
and attracts much interests recently \cite{guangyin_sci_rep_2013,mlynek_sup_rad_nat_com_2014}. In this system, the two atoms are coupled with each other via a common photon reservoir
with the wave number $q$. The master equation for the two-atom system is
\begin{eqnarray}
\frac{\partial}{\partial t}\rho(t)&=&\gamma\frac{\sin x}{x}\left(\hat{\sigma}_{1-}\rho\hat{\sigma}_{2+}-\frac{1}{2}\left\{\hat{\sigma}_{2+}\hat{\sigma}_{1-},\rho\right\}\right) \nonumber\\
&&+\gamma\frac{\sin x}{x}\left(\hat{\sigma}_{2-}\rho\hat{\sigma}_{1+}-\frac{1}{2}\left\{\hat{\sigma}_{1+}\hat{\sigma}_{2-},\rho\right\}\right) \nonumber\\
&&+\sum_{j=1,2}\gamma_j\left(\hat{\sigma}_{j-}\rho\hat{\sigma}_{j+}-\frac{1}{2}\left\{\hat{\sigma}_{j+}\hat{\sigma}_{j-},\rho\right\}\right),
\label{master_equation_exam4}
\end{eqnarray}
where $x\equiv qd$ and $d$ is the distance of the two atoms. Although there is no direct coupling between the two atoms, they can exchange energy via the common reservoir, as described
by the first two terms in the right-hand side of Eq.~(\ref{master_equation_exam4}).

It should be stressed that even if $X=(\sin x)/x$ can be negative for
certain $x$ and leads to an effective negative rate $\gamma X$, the dynamics
given by Eq.~(\ref{master_equation_exam4}) is, in fact, Markovian since the
Kossakowski matrix $\mathcal{K}$ \cite{gorini_lindbladform_jmp_1976}, which
is formed by collecting all the rates in front of each Lindblad term in Eq.~(%
\ref{master_equation_exam4}), is positive semidefinite. However, it is not
the case if we pay attention to only one atom among them by tracing out the
other one. The traced-out atom acts as a non-Markovian environment akin to
the C qubit in the second example. Let the the two atoms have no initial
correlation and the traced-out atom have an initial population $a$ on its
excited state. It can be viewed as the energy feedback to the system atom.

In Fig.~\ref{Fig_exam_4}, we show the divisibility phase diagram for the system atom. When $a=0$, the environment atom is in its ground state and no energy feedback into the system
atom can occur. Thus, the system atom undergoes a purely Markovian dissipation and the dynamics belongs to $\mathcal{PD}_2$. When $x$ is a multiple of $\pi$, both the atoms are effectively blind to each other due to the destructive interference. This inhibits the interatom energy exchange and also leads to the Markovian behavior of the system atom, regardless
of the value of $a$. In general, the system atom is more non-Markovian with increasing $a$, due to the stronger feedback of energy from the environment atom. Hence, $\mathcal{PD}_0$
distributes over the most of the region with large $a$, whereas $\mathcal{PD}_1$ can only occupy the region with small $a$ and the narrow regions surrounding $x=n\pi$.

In contrast to the previous examples, the BLP measure can only partially detect the non-Markovianity in the region of $\mathcal{PD}_0$, which is above the gray dashed curve in
Fig.~\ref{Fig_exam_4}. Furthermore, the LFS measure can even only work in the small corner above the green dashed curve. This is because the information from the environment to the
system atom is reduced when increasing the interatom distance $d$. However, the non-Markovianity should be kept even for a large distance, due to the interatom coupling mediated by
the reservoir photons \cite{guangyin_sci_rep_2013}. With the peculiar excitation-transfer rate in the form of ${\sin(qd)/qd}$, the transition between Markovian and non-Markovian
behavior should possess a periodicity of $qd=n\pi$, and the measure of non-Markovianity keeps finite but decreasing when increasing the distance.

\section{CONCLUSIONS}

In summary, we have utilized the theory of $k$-positivity maps to generalize the notion of CP divisibility to a hierarchy of $k$-divisibility. This results in a refinement of quantum
non-Markovianity and allows us to classify quantum processes into the partition of $\mathcal{D}_n$ in terms of $\mathcal{PD}_k$. Further visualization of the $\mathcal{PD}_{k}$ partition
leads to a useful tool, referred as $k$-divisibility phase diagrams. These phase diagrams show the landscape of non-Markovianity and allow one to study different measures in a unified
framework. We can acquire a deeper insight into the nature of quantum non-Markovianity, such as an intuitive way to realize the cause of the eternal non-Markovian dynamics, the reason
for the perfection of RHP measure, the clue for the weakness of BLP and LSF measures, and the circumstance when the measures get reconciled with each other.
Finally, we consider the superradiant phenomenon of a two-atom system which can be implemented experimentally.
The Markovian region is reduced to several straight lines instead of an area due to the destructive interference or the lack of feedback energy. The distribution of non-Markovian regions
is related to the rate of energy feedback to the system atom and possesses periodicity due to the peculiar form of photon-mediated interaction. In this case, the BLP measure can only
partially detect the non-Markovian in the $\mathcal{PD}_0$ and the LFS measure can even only work in a smaller corner. This weakness is due to that the increasing of the interatom
distance may reduce the back flow of information and thus the distinguishability can hardly increase with time.

$^\triangle$($\overline{\quad}_{(\infty)}\overline{\quad}$)$^\triangle$

\section*{ACKNOWLEDGMENTS}

This work is supported partially by the National Center for Theoretical
Sciences and Minister of Science and Technology, Taiwan, Grant No. NSC
101-2628-M-006-003-MY3 and No. MOST 103-2112-M-006-017-MY4. G.Y.C. is supported by
the Minister of Science and Technology, Taiwan, Grant No. MOST
102-2112-M-005-009-MY3.

\section*{APPENDIX}

\subsection{Details for the second example}

The second example considers a pair of qubits coupled with each other via a CNOT gate. As explained in the main text, the master equation for the pair of qubit can be expressed as
\begin{equation}
\frac{\partial}{\partial t}\rho_{\mathrm{CT}}(t)=-\frac{i}{\hbar}\left[\widehat{H}_{\mathrm{CT}},\rho_{\mathrm{CT}}\right]+
\sum_{j=1}^3\frac{\gamma_j}{2}\left(\hat{\sigma}_j\rho_{\mathrm{CT}}\hat{\sigma}_j-\rho_{\mathrm{CT}}\right),
\end{equation}
and the T qubit undergoes the isotropic depolarizing channel with
\begin{equation}
\gamma_1=\gamma_2=\gamma_3=\gamma.
\end{equation}

By solving the master equation and tracing out the C qubit, the dynamics of T qubit can be expressed as
\begin{equation}
\rho_\mathrm{T}(t)=\alpha_0\rho_{11}(t)+\beta_0\rho_{00}(t)+\delta_0\rho_{01}(t)+\delta^{\ast}_0\rho_{10}(t),
\end{equation}
where $\alpha_0$, $\beta_0$, $\delta_0$, and $\delta^{\ast}_0$ denote the initial condition of the T qubit and
\begin{widetext}
\begin{eqnarray}
\rho_{11}(t)&=&\left[\begin{array}{cc}
\frac{1}{2}\left(1+e^{-2\gamma t}\left(1-a+a\cos\frac{Jt}{\hbar}\right)\right) & i\frac{a}{2}e^{-2\gamma t}\sin\frac{Jt}{\hbar} \\
-i\frac{a}{2}e^{-2\gamma t}\sin\frac{Jt}{\hbar} & \frac{1}{2}\left(1-e^{-2\gamma t}\left(1-a+a\cos\frac{Jt}{\hbar}\right)\right)
\end{array}\right], \label{dynamics_11_example2}\\
\rho_{00}(t)&=&\left[\begin{array}{cc}
\frac{1}{2}\left(1-e^{-2\gamma t}\left(1-a+a\cos\frac{Jt}{\hbar}\right)\right) & -i\frac{a}{2}e^{-2\gamma t}\sin\frac{Jt}{\hbar} \\
i\frac{a}{2}e^{-2\gamma t}\sin\frac{Jt}{\hbar} & \frac{1}{2}\left(1+e^{-2\gamma t}\left(1-a+a\cos\frac{Jt}{\hbar}\right)\right)
\end{array}\right], \\
\rho_{01}(t)&=&\left[\begin{array}{cc}
-i\frac{a}{2}e^{-2\gamma t}\sin\frac{Jt}{\hbar} & ae^{-2\gamma t}\sin^2\frac{Jt}{2\hbar} \\
\frac{1}{2}e^{-2\gamma t}\left(2-a+a\cos\frac{Jt}{\hbar}\right) & i\frac{a}{2}e^{-2\gamma t}\sin\frac{Jt}{\hbar}
\end{array}\right], \\
\rho_{10}(t)&=&\left[\begin{array}{cc}
i\frac{a}{2}e^{-2\gamma t}\sin\frac{Jt}{\hbar} & \frac{1}{2}e^{-2\gamma t}\left(2-a+a\cos\frac{Jt}{\hbar}\right) \\
ae^{-2\gamma t}\sin^2\frac{Jt}{2\hbar} & -i\frac{a}{2}e^{-2\gamma t}\sin\frac{Jt}{\hbar}
\end{array}\right]. \label{dynamics_10_example2}
\end{eqnarray}
\end{widetext}

Having acquired the full dynamics of $\rho_\mathrm{T}(t)$, the corresponding Choi-Jamio{\l}kowski matrix can be constructed by combining
Eqs.~(\ref{dynamics_11_example2})-(\ref{dynamics_10_example2}) into the form of a block matrix,
\begin{equation}
\mathfrak{J}\left(\mathcal{E}_\mathrm{T}(t)\right)=\left[
\begin{array}{c|c}
\rho_{11}(t) & \rho_{10}(t) \\
\hline
\rho_{01}(t) & \rho_{00}(t)
\end{array}
\right].
\end{equation}
It is easy to find out that for any values of $J$, $\gamma$, and $a$, the four eigenvalues of $\mathfrak{J}\left(\mathcal{E}_\mathrm{T}(t)\right)$ are positive for any $t\geq 0$.
Namely, the dynamics of $\rho_\mathrm{T}(t)$ described by Eqs.~(\ref{dynamics_11_example2})-(\ref{dynamics_10_example2}) is CP for any $t\geq 0$.

Next, to determine the border of $\mathcal{PD}_2$ and $\mathcal{PD}_1$, we invert $\mathcal{E}_{\mathrm{T}}(t)$ and calculate the complement process
$\Lambda_{\mathrm{T}}(\tau,t)=\mathcal{E}_{\mathrm{T}}(t+\tau)\circ\left[\mathcal{E}_{\mathrm{T}}(t)\right]^{-1}$. The basic idea is outlined in
Ref.~\cite{andersson_kraus_decompo_jmo_2007}. One can rewrite the matrices in Eqs.~(\ref{dynamics_11_example2})-(\ref{dynamics_10_example2}) into the form of column vectors and combine
them together to construct $\mathcal{E}_\mathrm{T}(t)$ in terms of a linear map matrix
\begin{equation}
\mathfrak{L}\left(\mathcal{E}_\mathrm{T}(t)\right)=\left[
\begin{array}{c|c|c|c}
\tilde{\rho}_{11}(t) & \tilde{\rho}_{01}(t) & \tilde{\rho}_{10}(t) & \tilde{\rho}_{00}(t) \\
\end{array}
\right]
\end{equation}
The operation of $\mathcal{E}_\mathrm{T}(t)$ on an arbitrary initial state $\rho_\mathrm{T}(0)$ can be expressed as an usual matrix multiplication
$\tilde{\rho}_\mathrm{T}(t)=\mathfrak{L}\left(\mathcal{E}_\mathrm{T}(t)\right)\cdot\tilde{\rho}_\mathrm{T}(0)$. The tilde symbol above each $\rho$ reminds reader that these matrices are
now written in terms of column vectors.
The complement process in linear map matrix form is calculated by the matrix multiplication
\begin{eqnarray}
\mathfrak{L}\left(\Lambda_{\mathrm{T}}(\tau,t)\right)&=&\mathfrak{L}\left(\mathcal{E}_{\mathrm{T}}(t+\tau)\right)\cdot
\left[\mathfrak{L}\left(\mathcal{E}_{\mathrm{T}}(t)\right)\right]^{-1} \nonumber\\
&=&\left[
\begin{array}{c|c|c|c}
\tilde{\Lambda}_{11}(\tau,t) & \tilde{\Lambda}_{01}(\tau,t) & \tilde{\Lambda}_{10}(\tau,t) & \tilde{\Lambda}_{00}(\tau,t) \\
\end{array}\right]. \nonumber\\
\end{eqnarray}
Then the Choi-Jamio{\l}kowski matrix $\mathfrak{J}\left(\Lambda_{\mathrm{T}}(\tau,t)\right)$ can be constructed by rearranging $\mathfrak{L}\left(\Lambda_{\mathrm{T}}(\tau,t)\right)$ as
\begin{equation}
\mathfrak{J}\left(\Lambda_{\mathrm{T}}(\tau,t)\right)=
\left[\begin{array}{c|c}
\Lambda_{11}(\tau,t) & \Lambda_{10}(\tau,t) \\
\hline
\Lambda_{01}(\tau,t) & \Lambda_{00}(\tau,t)
\end{array}\right].
\end{equation}

By Theorem~\ref{theo_criterion_for_cp}, if the four eigenvalues of $\mathfrak{J}\left(\Lambda_{\mathrm{T}}(\tau,t)\right)$ are all positive for all $t,\tau\geq 0$, then
$\Lambda_{\mathrm{T}}(\tau,t)$ is CP for all $t,~\tau\geq 0$. Namely, the dynamics of $\rho_{\mathrm{T}}(t)$ described by Eqs.~(\ref{dynamics_11_example2})-(\ref{dynamics_10_example2})
is Markovian and belongs to $\mathcal{PD}_2$. On the other hand, if the CP property is violated for certain $t$ or $\tau$, then the dynamics belongs to $\mathcal{D}_1$. This
determines the border of $\mathcal{PD}_2$ and $\mathcal{PD}_1$.

Sequentially, to verify the positivity of $\Lambda_{\mathrm{T}}(\tau,t)$, we apply Proposition~\ref{prop_criteria_for_k_positivity_1} for $k=1$. Namely, applying
$\Lambda_{\mathrm{T}}(\tau,t)$ to an arbitrary state parameterized by $\{\theta,\phi\}$, one can obtain
\begin{eqnarray}
\Lambda_{\mathrm{T}}(\tau,t)\{\rho(\theta,\phi)\}&=&\cos^2\left(\frac{\theta}{2}\right)\Lambda_{11}(\tau,t) \nonumber\\
&&+e^{-i\phi}\sin\left(\frac{\theta}{2}\right)\cos\left(\frac{\theta}{2}\right)\Lambda_{01}(\tau,t) \nonumber\\
&&+e^{i\phi}\sin\left(\frac{\theta}{2}\right)\cos\left(\frac{\theta}{2}\right)\Lambda_{10}(\tau,t) \nonumber\\
&&+\sin^2\left(\frac{\theta}{2}\right)\Lambda_{00}(\tau,t).
\end{eqnarray}
It possesses two eigenvalues and here we denote them by $p_{\Lambda_{\mathrm{T}}}^{\pm}(\tau,t)$.
If the condition
\begin{equation}
0\leq p_{\Lambda_{\mathrm{T}}}^{\pm}(\tau,t)\leq 1,\quad\forall~t,\tau\geq 0
\label{criterion_exam2_in_PD1}
\end{equation}
is fulfilled, then $\Lambda_{\mathrm{T}}(\tau,t)$ is positive for all $t,~\tau\geq 0$. Namely, the dynamics of $\rho_\mathrm{T}$ belongs to $\mathcal{PD}_1$, otherwise
to $\mathcal{PD}_0$.

\subsection{Details for the third example}

The qubit dynamics subjected to amplitude-damping channel is governed by the master equation
\begin{equation}
\frac{\partial}{\partial t}\rho=\gamma(t)\left(\hat{\sigma}_{-}\rho\hat{\sigma}_{+}-\frac{1}{2}\left\{\hat{\sigma}_{+}\hat{\sigma}_{-},\rho\right\}\right).
\end{equation}
Its solution can be calculated analytically as
\begin{equation}
\rho_{\mathrm{AD}}(t)=\left[
\begin{array}{cc}
\rho_{++}(0)|G(t)|^2 & \rho_{+-}(0)G(t) \\
\rho_{-+}(0)G^\ast(t) & \rho_{++}(0)\left(1-|G(t)|^2\right)+\rho_{--}(0)
\end{array}
\right],
\end{equation}
where $|G(t)|^2=\exp\left[-\int_0^t\gamma(\tau)d\tau\right]$. The corresponding Choi-Jamio{\l}kowski matrix is written as
\begin{equation}
\mathfrak{J}\left(\mathcal{E}_{\mathrm{AD}}(t)\right)=\left[
\begin{array}{cccc}
|G(t)|^2 & 0 & 0 & G(t) \\
0 & 1-|G(t)|^2 & 0 & 0 \\
0 & 0 & 0 & 0 \\
G^\ast(t) & 0 & 0 & 1
\end{array}
\right].
\label{CJ_matrix_exam3}
\end{equation}
The four eigenvalues can be calculated easily as
\begin{equation}
\lambda_{\mathrm{AD}}(t)=1\pm|G(t)|,~0,~0.
\end{equation}
Hence, the condition for process $\mathcal{E}_{\mathrm{AD}}(t)$ being CP is
\begin{equation}
0\leq|G(t)|\leq 1,\quad\forall~t\geq 0.
\end{equation}

Next we calculate the complement process $\Lambda_{\mathrm{AD}}(\tau,t)$ and its Choi-Jamio{\l}kowski matrix
\begin{equation}
\mathfrak{J}\left(\Lambda_{\mathrm{AD}}(\tau,t)\right)=\left[
\begin{array}{cccc}
\frac{|G(t+\tau)|^2}{|G(t)|^2} & 0 & 0 & \frac{G(t+\tau)}{G(t)} \\
0 & 1-\frac{|G(t+\tau)|^2}{|G(t)|^2} & 0 & 0 \\
0 & 0 & 0 & 0 \\
\frac{G^\ast(t+\tau)}{G^\ast(t)} & 0 & 0 & 1
\end{array}
\right].
\label{CJ_matrix_complement_exam3}
\end{equation}
The four eigenvalues are
\begin{equation}
\lambda_{\Lambda_{\mathrm{AD}}}(\tau,t)=1\pm\frac{|G(t+\tau)|}{|G(t)|},~0,~0.
\end{equation}
Hence, the condition for process $\mathcal{E}_{\mathrm{AD}}(t)$ being in $\mathcal{PD}_2$ is
\begin{equation}
0\leq\frac{|G(t+\tau)|}{|G(t)|}\leq 1,\quad\forall~t,\tau\geq 0.
\end{equation}
This condition implies that $\mathcal{E}_{\mathrm{AD}}(t)\in\mathcal{PD}_2$ (being Markovian) if and only if $|G(t)|$
decreases monotonically in time.

We proceed to verify the positivity of $\Lambda_{\mathrm{AD}}(\tau,t)$. Applying Proposition~\ref{prop_criteria_for_k_positivity_1} for $k=1$, one can obtain
\begin{eqnarray}
&&\Lambda_{\mathrm{AD}}(\tau,t)\{\rho(\theta,\phi)\}= \nonumber\\
&&\left[
\begin{array}{cc}
\cos^2\frac{\theta}{2}\frac{|G(t+\tau)|^2}{|G(t)|^2} & e^{-i\phi}\cos\frac{\theta}{2}\sin\frac{\theta}{2}\frac{G(t+\tau)}{G(t)} \\
e^{-i\phi}\cos\frac{\theta}{2}\sin\frac{\theta}{2}\frac{G(t+\tau)}{G(t)} & 1-\cos^2\frac{\theta}{2}\frac{|G(t+\tau)|^2}{|G(t)|^2}
\end{array}
\right]. \nonumber\\
\end{eqnarray}
The two eigenvalues are
\begin{eqnarray}
&&p_{\mathrm{AD}}^{\pm}(\tau,t)= \nonumber\\
&&\frac{1}{2}\pm\sqrt{\frac{1}{4}+\frac{\left(1+\cos\theta\right)^2}{4}\left(\frac{|G(t+\tau)|^4}{|G(t)|^4}-\frac{|G(t+\tau)|^2}{|G(t)|^2}\right)}. \nonumber\\
\end{eqnarray}
After carefully analyzing the two eigenvalues, we can find out that the positivity of $p_{\mathrm{AD}}^{\pm}(\tau,t)$ implies that
$\mathcal{E}_{\mathrm{AD}}(t)\in\mathcal{PD}_1$ if and only if $|G(t)|$ decreases monotonically in time.

From the above discussions, one can draw the conclusion that the degeneracy occurs and only $\mathcal{PD}_2$ and $\mathcal{PD}_0$
leave in the $\mathcal{PD}_k$ partition.

Consider explicitly the Lorentzian spectral density function
\begin{equation}
J(\omega)=\frac{\gamma_0 \lambda^2}{\pi}\frac{1}{\omega^2+\lambda^2},
\label{spectral_density_function}
\end{equation}
where $\gamma_0$ represents the coupling constant between the system and environment, and $\lambda$ defines the spectral width. For the spectral density function in
Eq.~(\ref{spectral_density_function}), one has
\begin{eqnarray}
G(t)&=&\frac{1}{2}\left(1+\frac{\lambda}{\sqrt{\lambda^2-2\gamma_0\lambda}}\right)e^{-\frac{\lambda-\sqrt{\lambda^2-2\gamma_0\lambda}}{2}t} \nonumber\\
&&+\frac{1}{2}\left(1-\frac{\lambda}{\sqrt{\lambda^2-2\gamma_0\lambda}}\right)e^{-\frac{\lambda+\sqrt{\lambda^2-2\gamma_0\lambda}}{2}t}.
\end{eqnarray}
Then the monotonicity of $|G(t)|$ implies
\begin{equation}
\gamma_0\leq\frac{\lambda}{2}.
\label{criterion_exam3}
\end{equation}
Namely, if the condition in Eq.~(\ref{criterion_exam3}) is fulfilled, then the process $\mathcal{E}_{\mathrm{AD}(t)}$ belongs to $\mathcal{PD}_2$; otherwise it belongs to
$\mathcal{PD}_0$.

\subsection{Details for the fourth example}

From the two-atom dynamics given by Eq.~(\ref{master_equation_exam4}), the dynamics of the system atom can be described by the Choi-Jami{\l}kowski matrix
\begin{equation}
\mathfrak{J}\left(\mathcal{E}_\mathrm{SR}(t)\right)=\left[
\begin{array}{c|c}
\rho_{\mathrm{ee}}(t) & \rho_{\mathrm{eg}}(t) \\
\hline
\rho_{\mathrm{ge}}(t) & \rho_{\mathrm{gg}}(t)
\end{array}
\right],
\label{CJ_matrix_exam4}
\end{equation}
where
\begin{eqnarray}
\rho_{\mathrm{ee}}(t)&=&\left[
\begin{array}{cc}
aG_a(t)+G^2(t) & 0\\
0 & 1-\left[aG_a(t)+G^2(t)\right]
\end{array}\right], \nonumber\\
\rho_{\mathrm{eg}}(t)&=&\left[
\begin{array}{cc}
0 & aF_a(t)+G(t)\\
0 & 0
\end{array}\right], \nonumber\\
\rho_{\mathrm{ge}}(t)&=&\left[
\begin{array}{cc}
0 & 0\\
aF_a(t)+G(t) & 0
\end{array}\right], \nonumber\\
\rho_{\mathrm{gg}}(t)&=&\left[
\begin{array}{cc}
aH_a^2(t) & 0\\
0 & 1-aH_a^2(t)
\end{array}\right], \nonumber\\
G_a(t)&=&-\frac{2X^2}{1-X^2}e^{-2\gamma t}-\frac{1}{2}e^{-\gamma t} \nonumber\\
&&+\frac{1-3X}{4\left(1+X\right)}e^{-\gamma(1-X)t}+\frac{1+3X}{4\left(1-X\right)}e^{-\gamma(1+X)t},\nonumber\\
G(t)&=&e^{-\frac{\gamma}{2}(1-X)t}\cosh\left(\frac{\gamma}{2}Xt\right),\nonumber\\
H_a(t)&=&e^{-\frac{\gamma}{2}(1-X)t}\sinh\left(\frac{\gamma}{2}Xt\right),\nonumber\\
F_a(t)&=&X\left(e^{-\gamma t}-1\right)H_a(t).
\end{eqnarray}
After careful analysis, one can find out that the four eigenvalues of $\mathfrak{J}\left(\mathcal{E}_{\mathrm{SR}}(t)\right)$ are positive for all $x,~t\geq 0$ and $0\leq a\leq 1$. Namely,
$\mathcal{E}_{\mathrm{SR}}(t)$ is indeed a CP process. It is interesting to notice that if $a=0$, $\mathfrak{J}\left(\mathcal{E}_{\mathrm{SR}}\right)$ will reduce to
$\mathfrak{J}\left(\mathcal{E}_{\mathrm{AD}}\right)$ in Eq.~(\ref{CJ_matrix_exam3}). This is reasonable since, if $a=0$, the system atom can only dissipate its excitation energy
continuously and undergo an effective amplitude-damping process.

As shown in the previous examples, before determining the $\mathcal{PD}_k$ partition, we must calculate the complement process by inverting
$\mathcal{E}_{\mathrm{SR}}(t)$. The linear map form $\mathfrak{L}\left(\mathcal{E}_{\mathrm{SR}}(t)\right)$ can be constructed by rearranging
the elements of $\mathfrak{J}\left(\mathcal{E}_{\mathrm{SR}}(t)\right)$ in Eq.~(\ref{CJ_matrix_exam4}) as
\begin{equation}
\mathfrak{L}\left(\mathcal{E}_{\mathrm{SR}}(t)\right)=\left[
\begin{array}{c|c|c|c}
\tilde{\rho}_{\mathrm{ee}}(t) & \tilde{\rho}_{\mathrm{ge}}(t) & \tilde{\rho}_{\mathrm{eg}}(t) & \tilde{\rho}_{\mathrm{gg}}(t)
\end{array}
\right].\nonumber\\
\label{linear_map_exam4}
\end{equation}
Having expressed the process $\mathcal{E}_{\mathrm{SR}}(t)$ in terms of a linear map $\mathfrak{L}\left(\mathcal{E}_{\mathrm{SR}}(t)\right)$ in
Eq.~(\ref{linear_map_exam4}), the linear map matrix corresponding to the inverse process $\left[\mathcal{E}_{\mathrm{SR}}(t)\right]^{-1}$ is simply the inverse matrix $\left[\mathfrak{L}\left(\mathcal{E}_{\mathrm{SR}}(t)\right)\right]^{-1}$ of Eq.~(\ref{linear_map_exam4}). The one for complement process
$\Lambda_{\mathrm{SR}}(\tau,t)$ can be obtained by the usual matrix multiplication as
\begin{equation}
\mathfrak{L}\left(\Lambda_{\mathrm{SR}}(\tau,t)\right)=\mathfrak{L}\left(\mathcal{E}_{\mathrm{SR}}(\tau)\right)\cdot
\left[\mathfrak{L}\left(\mathcal{E}_{\mathrm{SR}}(t)\right)\right]^{-1}.
\label{linear_map_complement_exam4}
\end{equation}
Then the Choi-Jami{\l}kowski matrix $\mathfrak{J}\left(\Lambda_{\mathrm{SR}}(\tau,t)\right)$ can be obtained again by rearranging the elements of
linear map matrix $\mathfrak{L}\left(\Lambda_{\mathrm{SR}}(\tau,t)\right)$ in Eq.~(\ref{linear_map_complement_exam4}),
\begin{widetext}
\begin{equation}
\mathfrak{J}\left(\Lambda_{\mathrm{SR}}(\tau,t)\right)=\left[
\begin{array}{cccc}
\mathfrak{J}_{11} & 0 & 0 & \frac{aF_a(t+\tau)+G(t+\tau)}{aF_a(t)+G(t)} \\
0 & \mathfrak{J}_{22} & 0 & 0 \\
0 & 0 & \mathfrak{J}_{33} & 0 \\
\frac{aF_a(t+\tau)+G(t+\tau)}{aF_a(t)+G(t)} & 0 & 0 & \mathfrak{J}_{44}
\end{array}
\right],
\label{CJ_matrix_complement_exam4}
\end{equation}
where
\begin{eqnarray}
\mathfrak{J}_{11}&=&\frac{\left[aG_a(t+\tau)+G^2(t+\tau)\right]\left[1-aH_a^2(t)\right]-aH_a^2(t+\tau)\left[1-\left(aG_a(t)+G^2(t)\right)\right]}
{aG_a(t)+G^2(t)-aH_a^2(t)}, \nonumber\\
\mathfrak{J}_{22}&=&\frac{\left[1-\left(aG_a(t+\tau)+G^2(t+\tau)\right)\right]\left[1-aH_a^2(t)\right]-\left[1-aH_a^2(t+\tau)\right]\left[1-\left(aG_a(t)+G^2(t)\right)\right]}
{aG_a(t)+G^2(t)-aH_a^2(t)}, \nonumber\\
\mathfrak{J}_{33}&=&\frac{-\left[aG_a(t+\tau)+G^2(t+\tau)\right]aH_a^2(t)+aH_a^2(t+\tau)\left[aG_a(t)+G^2(t)\right]}
{aG_a(t)+G^2(t)-aH_a^2(t)}, \nonumber\\
\mathfrak{J}_{44}&=&\frac{-\left[1-\left(aG_a(t+\tau)+G^2(t+\tau)\right)\right]aH_a^2(t)+\left[1-aH_a^2(t+\tau)\right]\left[aG_a(t)+G^2(t)\right]}
{aG_a(t)+G^2(t)-aH_a^2(t)}.
\end{eqnarray}
\end{widetext}

As noticed in the main text, a special case occurs when $a=0$, i.e., the traced-out atom is initially in ground state. The dynamics reduced to the case of
amplitude-damping channel and the Choi-Jami{\l}kowski matrix in Eq.~(\ref{CJ_matrix_complement_exam4}) can be simplified as
\begin{equation}
\mathfrak{J}\left(\Lambda_{\mathrm{SR}}(\tau,t)\right)|_{a=0}=\left[
\begin{array}{cccc}
\frac{G^2(t+\tau)}{G^2(t)} & 0 & 0 & \frac{G(t+\tau)}{G(t)} \\
0 & 1-\frac{G^2(t+\tau)}{G^2(t)} & 0 & 0 \\
0 & 0 & 0 & 0 \\
\frac{G(t+\tau)}{G(t)} & 0 & 0 & 1
\end{array}
\right],
\end{equation}
apparently the same form as that in Eq.~(\ref{CJ_matrix_complement_exam3}). From the conclusion in the previous section,
$\mathcal{E}_{\mathrm{SR}}(t)$ is in $\mathcal{PD}_2$ for $a=0$.

Another special case occurs as $x=n\pi$, where the factor $X=0$. Destructive interference takes place and forbids the energy exchange between the two atoms. Hence, the dynamics of
the system atom is independent of the initial state of the traced-out atom and the Choi-Jami{\l}kowski matrix is again reduced into the form of Eq.~(\ref{CJ_matrix_complement_exam3}),
\begin{equation}
\mathfrak{J}\left(\Lambda_{\mathrm{SR}}(\tau,t)\right)|_{x=n\pi}=\left[
\begin{array}{cccc}
\frac{G^{\prime 2}(t+\tau)}{G^{\prime 2}(t)} & 0 & 0 & \frac{G^{\prime}(t+\tau)}{G^{\prime}(t)} \\
0 & 1-\frac{G^{\prime 2}(t+\tau)}{G^{\prime 2}(t)} & 0 & 0 \\
0 & 0 & 0 & 0 \\
\frac{G^{\prime}(t+\tau)}{G^{\prime}(t)} & 0 & 0 & 1
\end{array}
\right],
\end{equation}
with $G^{\prime}(t)=G(t)|_{x=n\pi}=\exp\left[-\gamma t/2\right]$. Similarly, from the conclusion in the previous section, $\mathcal{E}_{\mathrm{SR}}(t)$ is in $\mathcal{PD}_2$ for
$x=n\pi$. Besides these two special cases, the eigenvalues of $\mathfrak{J}\left(\Lambda_{\mathrm{SR}}(\tau,t)\right)$ are quite complicated. After numerical analysis, we find out
that $\mathfrak{J}\left(\Lambda_{\mathrm{SR}}(\tau,t)\right)$ fails to be positive semidefinite. The process $\mathcal{E}_{\mathrm{SR}}(t)$ is non-Markovian besides the two special
cases.

Next, to verify the positivity of $\Lambda_{\mathrm{SR}}(\tau,t)$, we apply it to an arbitrary state parameterized by $\{\theta,\phi\}$ and investigate
the positivity of $\Lambda_{\mathrm{SR}}(\tau,t)\{\rho(\theta,\phi)\}$. Due to the complexity, one can only invoke numerical analysis. If the positivity
of $\Lambda_{\mathrm{SR}}(\tau,t)\{\rho(\theta,\phi)\}$ holds for all $t$, $\tau$, $\theta$, and $\phi$, then the process $\mathcal{E}_{\mathrm{SR}}(t)$
is in $\mathcal{PD}_1$. These lead to the $k$-divisibility phase diagram in Fig.~4 in the main text.


%

\end{document}